\documentclass[runningheads]{llncs}
\usepackage{graphicx}
\usepackage{aas_macros}
\usepackage{amssymb}
\usepackage{tabularx}
\usepackage{hyperref}

\newcommand{\code}[1]{\texttt{#1}}

\begin{document}
\title{Autonomous real-time science-driven follow-up of survey transients}
%
%
\author{Niharika Sravan\inst{1} \and
Matthew J. Graham\inst{1} \and
Christoffer Fremling\inst{1} \and
Michael W. Coughlin\inst{2}}
\authorrunning{N. Sravan et al.}
%
\institute{Division of Physics, Mathematics, and Astronomy, California Institute of Technology, Pasadena, CA 91125, USA \and
School of Physics and Astronomy, University of Minnesota, Minneapolis, Minnesota 55455, USA}
\maketitle              
\begin{abstract}
Astronomical surveys continue to provide unprecedented insights into the time-variable Universe and will remain the source of groundbreaking discoveries for years to come. However, their data throughput has overwhelmed the ability to manually synthesize alerts for devising and coordinating necessary follow-up with limited resources. The advent of Rubin Observatory, with alert volumes an order of magnitude higher at otherwise sparse cadence, presents an urgent need to overhaul existing human-centered protocols in favor of machine-directed infrastructure for conducting science inference and optimally planning expensive follow-up observations.

We present the first implementation of autonomous real-time science-driven follow-up using value iteration to perform sequential experiment design. We demonstrate it for strategizing photometric augmentation of Zwicky Transient Facility Type Ia supernova light-curves given the goal of minimizing SALT2 parameter uncertainties. We find a median improvement of 2-6\% for SALT2 parameters and 3-11\% for photometric redshift with 2-7 additional data points in $g$, $r$ and/or $i$ compared to random augmentation. The augmentations are automatically strategized to complete gaps and for resolving phases with high constraining power (e.g. around peaks). We suggest that such a technique can deliver higher impact during the era of Rubin Observatory for precision cosmology at high redshift and can serve as the foundation for the development of general-purpose resource allocation systems.
\end{abstract}
\section{Introduction}
Over the past two decades, increasingly sophisticated all-sky surveys have ushered in a golden era of time-domain astronomy. Not only has it produced unprecedented statistics \cite{2017ApJ...837..121G,2019ApJ...874..106B,2020ApJ...895...32F} and 
revealed an astonishing diversity of transients \cite{2016ApJ...819...35A,2017ApJ...836...60L,2019ApJ...872...18M}, but has paved the way for putting together a rich understanding of the physical processes governing our Universe. More recently, our ability to probe transients synchronously in electromagnetic and gravitational waves have enabled inquiries in regimes previously out of reach \cite{2019MNRAS.489L..91C,2021arXiv210211569R,2021arXiv210706229H}. The advent of the Rubin Observatory in the upcoming decade will transform this paradigm by increasing the discovery rate by at least an order of magnitude over the current rate \cite{2019ApJ...873..111I}. However, several limitations prevent us from fully exploiting the science potential of even current survey throughput \cite{2016arXiv160104385D,2018ApJS..236....9N,2019NatRP...1..600H}. 

The primary issue is that survey data volumes exceed the capacity of human experts to process.  Moreover, in many cases, additional follow-up is necessary to allow detailed science analyses (e.g. obtaining spectra to estimate redshift or multi-wavelength observations to probe circumstellar medium interaction). However, even with a curated list, the number of candidates for follow-up far exceed the availability of suitable resources. For transients, the situation is further exacerbated by the time-sensitivity of decision-making for such allocation while assessing trade-offs between competing science objectives. Unsurprisingly, this situation results in inefficiencies and lost opportunities.

Significant progress has been made in response to these issues. These include data broker systems (e.g., ALeRCE \cite{2021AJ....161..242F}; AMPEL \cite{2019A&A...631A.147N}; Antares \cite{2014SPIE.9149E..08S}; Fink \cite{2021MNRAS.501.3272M}; Fritz\footnote{https://github.com/fritz-marshal/fritz}, Lasair \cite{2019RNAAS...3a..26S}; MARS\footnote{\href{https://mars.lco.global/}{https://mars.lco.global/}}; Pitt-Google Broker\footnote{\href{https://pitt-broker.readthedocs.io/en/latest/}{https://pitt-broker.readthedocs.io/en/latest/}}), that sort, value-add, and curate alert streams for downstream science groups, and Target and Observation Managers or marshals  \cite{2018SPIE10707E..11S,2019PASP..131c8003K}, that help them plan and co-ordinate follow-up. In addition, robust and efficient classifiers for complete \cite{2019AJ....158..257B,2021AJ....161..141S}, partial \cite{2019PASP..131k8002M,2020MNRAS.491.4277M,2020arXiv200803309C}, or no light-curves \cite{2020ApJ...902...60B} have been developed to aid in follow-up decision-making. Tools to aid in discovery, such as those for identifying out-of-distribution events, have also been developed \cite{2021MNRAS.502.5147M,2021ApJS..255...24V,2021arXiv211100036M,2021A&C....3600481L}. 

An important but relatively underdeveloped component is systems that autonomously strategize optimal follow-up for an arbitrary set of science goals. Designated ORACLEs\footnote{Object Recommender for Augmentation and Coordinating Liaison Engine} \cite{2020ApJ...893..127S}, they share a lot of synergy with brokers and marshals and are an important component of the overall software infrastructure needed to support survey science. Such systems are similar to those for selecting objects to obtain spectra with the goal of overcoming Malmquist bias in training data \cite{2012ApJ...744..192R,2020arXiv201005941K,2020AJ....159...16A} except that the prioritization is based on science utility. 

While optimal follow-up strategies have been studied for populations of events to determine general guiding principles \cite{2019ApJ...880L..22W,2020ApJ...889...36C}, ideally these would be determined on a case-by-case basis and be adaptive to accommodate additional information that becomes available.  Recent work \cite{2021arXiv210609761C} solved the optimal resource allocation problem for obtaining galaxy spectra (to derive distances and masses), constrained by an observing budget, to maximize Shannon Information on $\Omega_m$ over all measurements. In this work, we solve the problem of optimal photometric follow-up of transient light-curves (LCs) to maximize constraints on theoretical LC models. It differs from \cite{2021arXiv210609761C} in that the design problem is sequential and executed in real-time. We are motivated to explore photometric augmentation by Rubin Observatory's Legacy Survey of Space and Time (LSST), $\sim$90\% of which will consist of the Wide Fast Deep (WFD) survey with a nominal cadence of $\sim$3 days in any filter ($u$, $g$, $r$, $i$, $z$, or $y$) and $\sim$15 days in the same filter. It is unclear to what extent the resulting sparse LCs could be directly interpreted with theoretical models. This is especially true for fast-evolving transients, including many core-collapse supernovae (SNe). Value-driven augmentation would maximize science uniquely possible due to LSST especially at the fainter end. 

In this paper, we focus on optimally augmenting SN Ia LCs from Phase-I of the Zwicky Transient Facility (ZTF) \cite{2019PASP..131a8002B,2019PASP..131g8001G} public survey for the goal of maximizing constraints on SALT2 model parameters \cite{2014A&A...568A..22B}. However, the framework is generalizable to any transient type and set of theoretical models. In Section \ref{s:meth_alg} we present the problem statement and our solution for strategizing real-time follow-up under uncertainty. We present results from our method in Section \ref{s:results} and conclude in Section \ref{s:conclusions}.

We assume a flat $\Lambda$CDM cosmology with $H_0=$ 70km/s/Mpc, $\Omega_m=1-\Omega_\Lambda$ = 0.3.

\section{Problem Statement and Algorithms} \label{s:meth_alg}
We aim to augment ZTF Phase-I branch-normal SN Ia LCs in $g$ and $r$ with photometry in $g$, $r$, and $i$ using the same instrument to maximize constraints on SALT2 models \cite{2007A&A...466...11G}. SALT2 models are used widely in cosmological analyses to measure distances using the Phillips' relation \cite{1993ApJ...413L.105P}. They are described by three parameters, $x_0$, $x_1$, and $c$, where $x_0$ is a scaling parameter, $x_1$ is a stretch parameter, $c$ is a color term. During Phase-I, ZTF conducted a public survey of the visible Northern sky every 3 days and the visible Galactic plane every day in $g$ and $r$ using a $\sim 47$ deg$^2$ imager on the 48-inch telescope on Palomar. We include $i$-band augmentation in our analysis to ascertain the importance of follow-up in different regimes. As such, data in redder filters are also important for deriving physics, including understanding the explosion mechanism \cite{2010AJ....139..120F} and estimating $H_0$ \cite{2018ApJ...869...56B}. 
To ensure we can accurately model the LC in $i$-band we select SNe Ia for which there is at least one $i$-band data within 60 days of trigger.   
SNe Ia used here were classified by the Bright Transient Survey (BTS), which aims to conduct a complete census of $r\lesssim$ 18.5 mag transients \cite{2020ApJ...895...32F,2020ApJ...904...35P} using a low resolution IFUS on the Palomar 60-inch telescope \cite{2019A&A...627A.115R}. 

\subsection{Optimal Real-time Decision under Uncertainty} \label{ss:big}
The problem of strategizing optimal follow-up in real-time can be thought of as follows. We are sequentially presented with available actions to choose from ($\mathcal A$, e.g. observe in a certain passband), each associated with some cost (e.g. exposure time), that map to a distribution of outcome states ($\mathcal S$, dictated by nature e.g. observed magnitude) and some utility (dictated by our objective, e.g. improvement in model constraints). In other words, we must assess the explore-exploit tradeoff between available actions given their expected utility, we cannot undo past actions but should adapt to new information, and there is some uncertainty about the future. Moreover, we may expect some information to become available in the future (e.g. if a survey is scheduled to revisit a field in some interval). The goal is to find the optimal set of actions, constrained by a budget, that maximizes the net utility of outcome states.

The solution is to choose the action with the maximum expected utility at every timestep, where the expected utility of action $\mathcal A$, $EU(\mathcal A)$, is given by:
\begin{equation} \label{e:eu}
    EU(\mathcal A) = \int\displaylimits_{\mathcal S} dP_{\mathcal A}(\mathcal S)U(\mathcal S) 
\end{equation}
where $P_{\mathcal A} (\mathcal S)$ is the probability of outcome state $\mathcal S$ for an action $\mathcal A$ and $U(\mathcal S)$ is the utility of state $\mathcal S$. The choice of utility conveys our preference for a given outcome state. Popular choices are A- or D-optimal designs, that minimize the average variance or maximize the Shannon Information of model parameters, respectively. However, more sophisticated utility functions, to aid in model discrimination \cite{2020ApJ...889...36C}, min-maxing, among others, also exist. In this work we use a proxy for A-optimality as our choice for utility (see Section \ref{ss:subprob}). We assume a discrete action space (one visit per night in a given filter), a cost of one per action, and a fixed budget per SN Ia.

In practice, it can be computationally difficult to evaluate $EU(\mathcal A)$ for all actions at a given timestep (i.e. all possible combinations of follow-up at the current and all future epochs). It can also be difficult to compute the utility (e.g. computing the Information matrix using sampling), especially for continuous outcome states. It is therefore useful to bound the problem and/or approximate the utility function, as in a deep Q-network \cite{2015Natur.518..529M}. We discuss our simplifications for the problem of augmenting ZTF SN Ia LCs to minimize SALT2 uncertainties next. 

\subsection{Sequential `Pseudo A-optimal' Experiment Design} \label{ss:subprob}
First, instead of finding the optimal action set for the full budget across the entire episode, we find the optimal set at the current epoch, i.e.
\begin{equation} 
\mathcal A = \{\emptyset, g, r, i, gr, ri, ig, gri\} 
\end{equation} \label{e:act}
and allocate the remaining budget randomly across future epochs. This substitutes future optimal actions for future mean actions.
Second, we assume a fixed outcome state per action, i.e.
\begin{equation} 
dP_{\mathcal A}(\mathcal S) = \delta(\mathcal S_{\mathcal A}^*-S) d_{\mathcal A}(\mathcal S)
\end{equation} \label{e:prob}
where $\mathcal S^*$ is outcome state associated with $\mathcal A$, i.e. $S_{\mathcal A}^*= \{{\rm photometry}(\mathcal A),$ ${\rm remaining~budget~random}, {\rm observed~photometry}, \rm {future~survey~photometry}\}$ and $\delta$ is the Kronecker delta function. We adopt the inverse mean SALT2 parameter uncertainty as our utility for $\mathcal S$. This is meant to serve as a proxy for A-optimality, since \code{sncosmo} solves the $\chi^2$ minimization problem, which is not equal to a maximum likelihood estimate. Fitting SALT2 models requires an estimate of the source redshift since the LCs are fit in rest-frame. Since this may not be available in real-time settings, we solve for the photometric redshift along with SALT2 parameters. We do, however, apply Milky-way extinction \cite{1989ApJ...345..245C,1998ApJ...500..525S} given the known SN sky location. Since, in many cases, $EU(\mathcal A)$ differ numerically by only a very small value, we require an improvement of $\epsilon$ over the expected utility of no action to take that action. A smaller value of $\epsilon$ results in greedy actions and spending the budget early. We tune the hyperparameter $\epsilon$ given a budget to maximize $EU(\mathcal A)$ for all ZTF SNe Ia. 

Every night, we estimate the $EU(\mathcal A)$ using Equations \ref{e:act} and \ref{e:prob}. This requires estimating $S_{\mathcal A}^*$. In real-time applications, this can be estimated using machine-learning based forecasting. In this paper, we focus on bracketting the upper limit performance of our framework by using a hypothetical forecaster (we report on real-time performance using an encoder-decoder LSTM in an upcoming work; Sravan et al., in prep). We use 2-D Gaussian Process regression on the full observed LC in $g$, $r$, and $i$ (as discussed in \cite{2020ApJ...893..127S} with minor modifications to operate in magnitude space) as our hypothetical forecaster. Estimates using this yield a realistic but upper limit performance since it leverages future photometry to estimate the outcome for the present state. We do not use SALT2 or any SN Ia LC models based on the Phillips relation \cite{1993ApJ...413L.105P} for this purpose because our science objective is intended to maximize constraints on the same and doing so would feed back any model errors or biases \cite{2013A&A...560A..66R,2014ApJ...784...51K}. Given mean magnitudes from our Gaussian Process regressor, we simulate magnitude uncertainties using a skew normal fit to BTS SN magnitude uncertainties in each passband and magnitude bin\footnote{bin edges at 12, 15.5, 17.5, 19.5, and 21.5 mag}. We do not account for magnitude uncertainty correlations on a given night. 

Estimating $S_{\mathcal A}^*$ also requires estimating future survey behaviour. For this we simulate an observing strategy for ZTF Phase-I in $g$ and $r$ by drawing revisit cadences from a Kernel Density Estimate (KDE)\footnote{Gaussian kernel with bandwidth of 0.005} for revisit intervals for BTS SNe. We do not account for gaps in cadence due to the moon or systematic difference in cadence for brighter/fainter SNe. For every revisit we assign a random night's observing strategy in a random BTS SNe. This assumes independence of observing strategies per visit which we verify using a $\chi^2$-contingency test. Our test yielded a p-value of 0.92 and we fail to accept our null hypothesis, that two consecutive nights' observing strategy, defined as a sequence of observed passbands, are dependent. We then estimate photometry at the simulated observing strategy using the method described above.

\begin{figure*}
\begin{center}
\includegraphics[width=\textwidth]{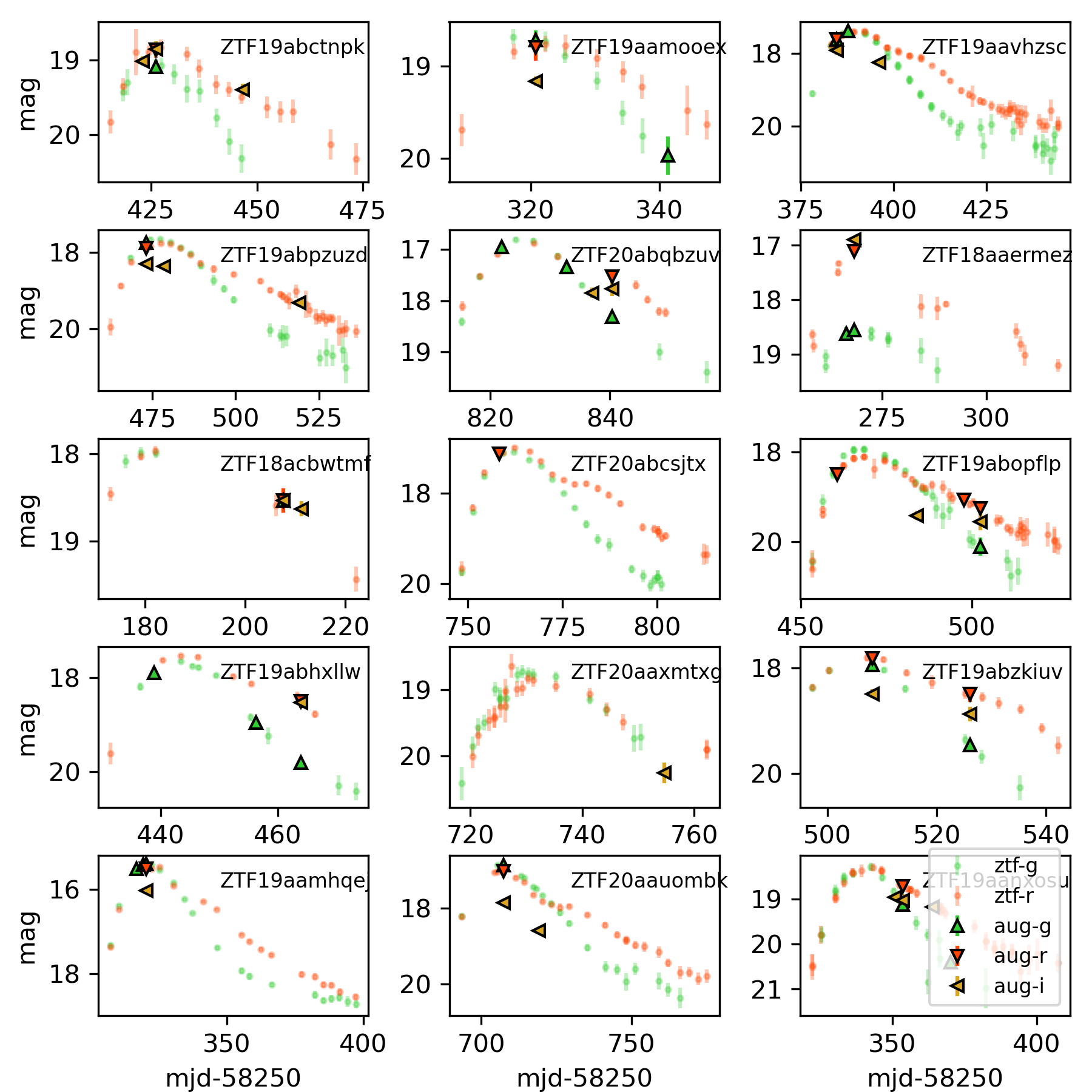}
\end{center}
\caption{Random sample of ZTF SN Ia LCs augmented in real-time given the goal of minimizing SALT2 parameter and photometric redshift uncertainties. Translucent green and red error-bars with circular markers are ZTF difference imaging photometry in $g$ and $r$, respectively. Green (upright), red (downwards), and yellow (left-pointing) triangular markers are simulated augmented photometry in $g$, $r$, and $i$.
The augmentations are typically strategized around first/second peaks and valleys and to complete gaps in LCs. 
\label{f:example}}
\end{figure*}

Finally, we take the action with the maximum expected utility and repeat the next day. To ensure robustness to stochastic outcomes for future survey sampling (described above), we simulate the $S_{\mathcal A}^*$ ten times and take the modal action with maximum $EU(\mathcal A)$ across all simulations. We do not augment if ZTF observed that day from the true observed LC to avoid duplicate observations. To simulate observed photometry given a chosen action, we use use the method described earlier for estimating $S_{\mathcal A}^*$. The simulated observed photometry are made available to determine the observing strategy for the next day. 

\section{Science Gain from Real-time Photometric Augmentation of ZTF SNe Ia} \label{s:results}

\begin{table} 
\centering
\caption{Median (IQR) improvement for SALT2 parameter ($x_0$, $x_1$, and $c$) and photometric redshift ($z$) uncertainty over random augmentation of SNe Ia LCs in $g$, $r$, and $i$.}
\label{t:perf}
\begin{tabularx}{\textwidth}{X X X X X X X X} 
\hline
Budget & $\delta \sigma_{x_0}$ & $\delta \sigma_{x_1}$ & $\delta \sigma_{c}$ & $\delta \sigma_{z}$ & $\epsilon$ & Expense & Frac Aug\\
\hline
3 & 0.02 (0.21) & 0.03 (0.18) & 0.05 (0.21) & 0.06 (0.51) & 0.78 & 2 & 0.34\\
6 & 0.03 (0.24) & 0.05 (0.22) & 0.04 (0.22) & 0.06 (0.71) & 0.92 & 5 & 0.11\\
9 & 0.05 (0.24) & 0.06 (0.20) & 0.05 (0.22) & 0.11 (0.64) & 0.96 & 7 & 0.08\\
\hline
\end{tabularx}
{\raggedright {\bf Notes -} \par}
{\raggedright Expense: Median utilization of budget \par}
{\raggedright Frac aug: Fraction of events that received at least one augmentation \par}
\end{table}

Table \ref{t:perf} shows the improvement for SALT2 parameter and photometric redshift uncertainty over random augmentation with the same expenditure. Note that not entire the budget is exhausted for all SNe Ia. However, of 497 SNe Ia $\sim$70-90\% received at least one augmentation. As expected, $\epsilon$ decreases with budget. This is because the threshold of taking an action is lower if more budget is available. Figure \ref{f:example} shows example ZTF SNe Ia augmented using our framework given a budget of six. Typically, augmentations are strategized to fill gaps, and around LC peaks and valleys. While we do not assume a spectroscopic redshift while strategizing follow-up, if spectroscopic redshift is eventually used (using either the SN or host spectrum) to fit the LCs the improvement decreases by $\sim$2\%. While the associated mean values for $x_0$ and $z$ are not affected much, those for $x_1$ and $c$ change by 4-8\% and 8\%, respectively. We leave the exploration of the impact on cosmological analyses to upcoming work.

\begin{figure*}
\begin{center}
\includegraphics[width=0.7\textwidth]{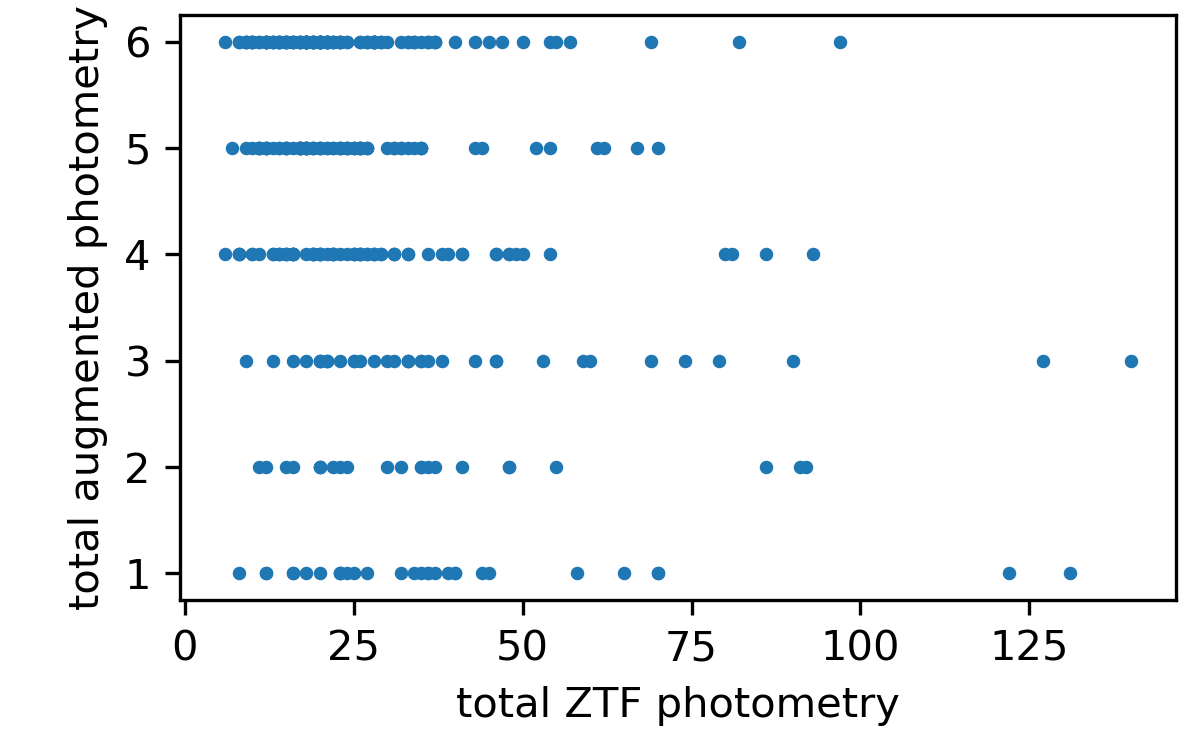}
\end{center}
\caption{Total number of augmented vs ZTF public survey photometry. Augmentations are strategized to fill in sparse LCs.
\label{f:aug-den}}
\end{figure*} 

\begin{figure*}
\begin{center}
\includegraphics[width=0.9\textwidth]{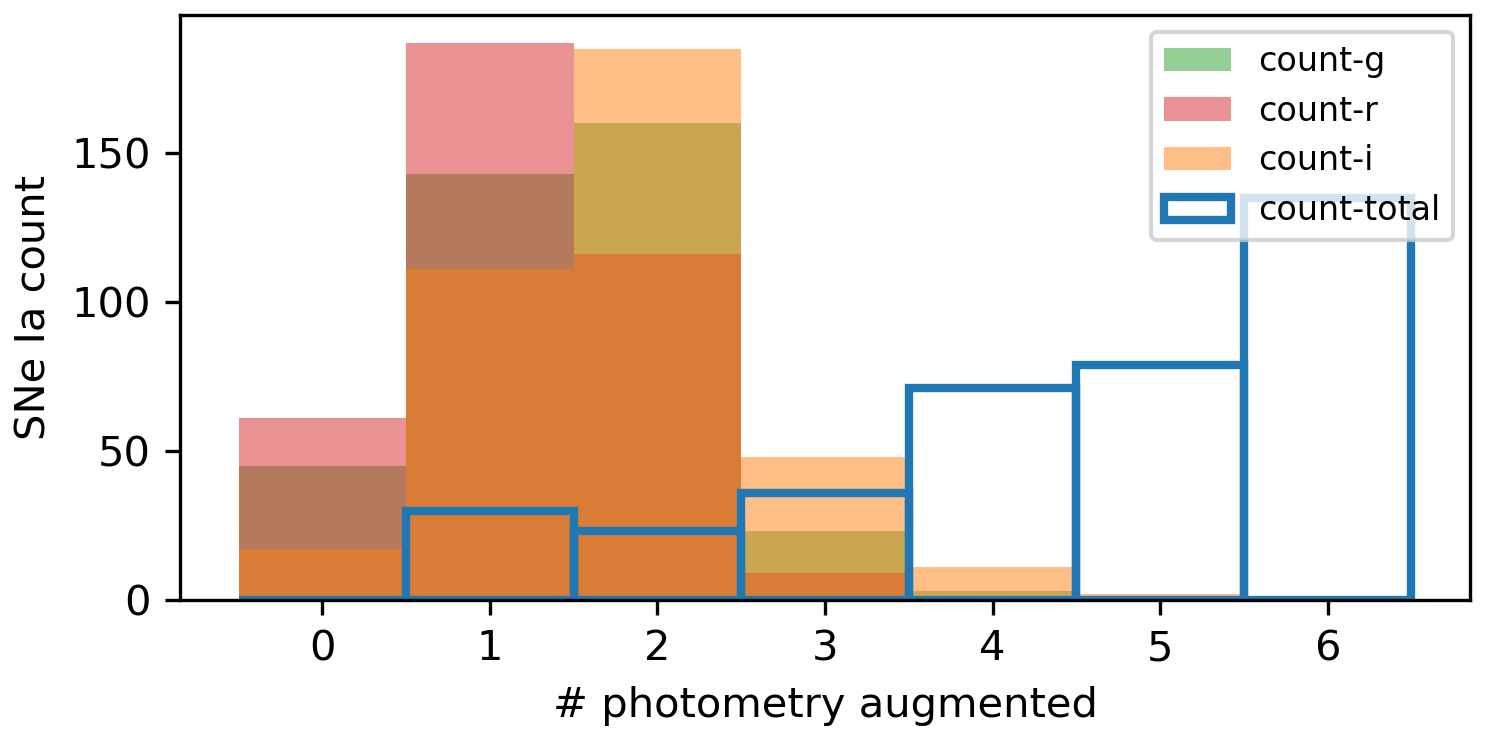}
\end{center}
\caption{Distribution of number of photometry augmented. There are slightly more augmentations in $i$, than $g$, followed by $r$.
\label{f:naug}}
\end{figure*} 

\begin{figure*}
\begin{center}
\includegraphics[width=0.9\textwidth]{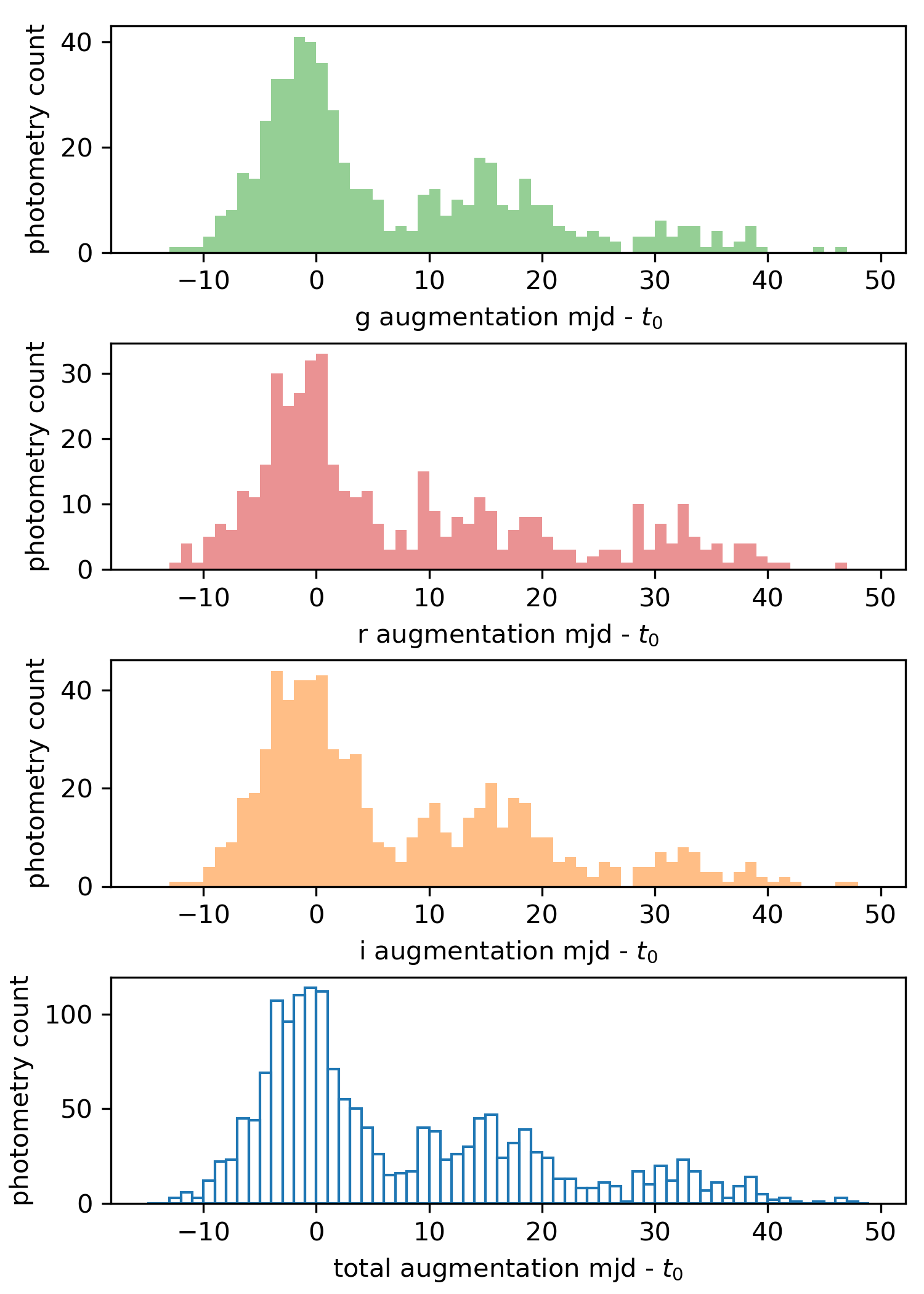}
\end{center}
\caption{Distribution of time to/since peak ($t_0$) at which photometry is augmented (budget=6). Augmentations are typically strategized around first and second peaks/valleys.
\label{f:phase-aug}}
\end{figure*} 

Figure \ref{f:aug-den} shows the number of augmentations vs sparsity of the ZTF survey LCs given a budget of six. As mentioned earlier, sparse LCs result more augmentations. 
Sparser LCs also experience more improvement. For instance, events with less than an average of one photometry every three days points (similar to that of LSST WFD) experience 2-5\% more improvement than random allocation (7\%, 7\&, 9\%, and 10\% for $\sigma_{x_0}$, $\sigma_{x_1}$, $\sigma_{c}$, and $\sigma_{z}$ for a budget of 6). This suggests greater returns from such an approach for LSST-like LCs. 
Figure \ref{f:naug} shows the distribution of the number of augmented photometry. Most augmentations are strategized in the $i$-band followed by $g$, then $r$. This is because the second peak of a SN Ia LC is most pronounced in the $i$-band. On the other hand, even though the second peak is more prominent in $r$ than $g$, since $g$-band LCs are typically fainter than $r$, it has sparser coverage and thus receives more augmentations. 

Figure \ref{f:phase-aug} shows the distribution of phase at which augmentations are strategized. Here $t_0$ is time of the bolometric LC peak from \code{sncosmo} fits to ZTF public survey LCs.  Half the augmentations are solicited within 2 days of peak and 90\% of are strategized within 26 days. The distributions have three modes that become more pronounced at redder wavelengths. This is once again because augmentations are strategized around LC phase with high varibility, i.e. around peaks and valleys.

\section{Conclusions} \label{s:conclusions}
In this paper, we solve the problem of real-time optimal resource allocation and demonstrate it for the case of photometric follow-up of ZTF SNe Ia given the goal of minimizing mean SALT2 parameter and photometric redshift uncertainty. We use the framework of optimal decision under uncertainty and sequential experiment design to achieve this. The framework produces a small but robust improvement for target parameters with reasonable expenditure. It automatically strategizes follow-up to improve LC sparsity and for resolving phases with high variability. Though SNe Ia analyzed here are low redshift ($\lesssim 0.15$), such an approach could yield important constraints for cosmology if applied to photometrically classified survey LCs at high redshift, especially in conjunction with programs such as RESSPECT \cite{2020arXiv201005941K}. We analyze the realistic performance of our framework using an encoder-decoder LSTM and the impact on cosmological inference, e.g. $H_0$ or the Hubble residual, in upcoming work.

There are a few modifications to our approach that are necessary to allow tackling of diverse and more complex real-world scenarios. First is expanding the follow-up action space to include more wavelengths. This would require increasing the cardinality of the action space to $\sum_{i=0}^N {N\choose i}$, where N is the total number of follow-up regimes (e.g. filters). Second is strategizing optimal follow-up given several concurrent candidate events. If the utility of the action space of each event is independent and there are enough resources to follow-up all of them, then the existing framework can simply be used to strategize follow-up for each of them separately. However, for a joint utility function (e.g. if follow-up for one event provides information about another), the action space would need to be increased to evaluate all possible follow-up scenarios for all events simultaneously. Related to the issue of limited follow-up resources, third, is accounting for a variable cost of follow-up (e.g. a function of integration time), a observing season based budget, and any data acquisition latency. We note that our framework is tolerant to data acquisition failures, i.e. the framework proceeds the next day normally, ignoring the previous night's strategy. Finally, while in this work we focus on optimal follow-up for improving parameter constraints, such an approach can be used to design long-term follow-up campaigns, for e.g. building priors on physical parameters or discriminating between theoretical models for populations of events.

In this work, we focus on SNe Ia because these are quite abundant, relatively well-defined, with robust theoretical models and public modeling infrastructure supporting it. However, there are potentially other science use cases that are more suited to this type of method. Typically, these would be for LCs that are either sparse (e.g. those from LSST WFD) and/or fast-evolving. One important application is the potential to inform optimal follow-up of gravitational-wave kilonova counterparts \cite{2019ApJ...885L..19C,2020A&A...643A.113A}, constrain $H_0$ \cite{2017Natur.551...85A,2019NatAs...3..940H,2020Sci...370.1450D} or discriminate between binary neutron star and neutron star-neutron star LC models \cite{2019PhRvD.100d3011C,2019PhRvD.100f3021H,2021NatAs...5...46A}.
Given the high survey data volumes, limited expert time and follow-up resources, delegating autonomous systems to handle data acquisition would free up time to not only further improve the science return for the particular field being targeted, but also for other science cases outside the purview of machine intelligence.

%
%
%
\bibliographystyle{splncs04}
\bibliography{references-master}

\end{document}